\documentclass[fleqn,twoside]{article}
\usepackage{espcrc2}

\usepackage{graphicx}
\usepackage[figuresright]{rotating}

\newcommand{\AmS}{{\protect\the\textfont2
  A\kern-.1667em\lower.5ex\hbox{M}\kern-.125emS}}

\title{Chiral Dirac fermions on the lattice using Geometric Discretisation {\thanks{Talk presented by V. de Beauc\'e.}}}
\author{V. de Beauc\'e \address{School of Mathematics, 
        University of Dublin, Trinity College, \\ 
        Dublin 2, Ireland}%
        \thanks{Supported by the Higher Education Authority of Ireland through the IITAC.}
        , S. Sen\addressmark \hspace{0.5mm} 
and 
J. C. Sexton\addressmark}

\begin{document}

\begin{abstract}
A new approach to the problem of doubling \cite{us} is presented with the Dirac-K\"ahler (DK) theory as a starting point and using Geometric Discretisation \cite{Us} providing us with a new way of extracting the Dirac field in the discrete setting of a hyper-cubic complex. 
\vspace{1pc}
\end{abstract}

\maketitle

\section{Introduction}

The theorem of Nielsen and Ninomiya \cite{NN} came with a topological proof of the fact that under reasonable assumptions, fermion doubling is unavoidable on the lattice. A key element of the proof was the periodicity of the Brillouin zone thus setting the approach in momentum space. 

In other arguments \cite{rabin}, \cite{BJ} it was argued that doubling is already present when one starts one step back and consider the DK equation (\ref{dke2})
and it persists after reduction to the Dirac equation on the lattice. The discretised field is then an inhomogeneous cochain taking value on points, edges and so on and it is a 16D object in four dimensions. The DK equation is also particularly well suited for counting fields through its link with the Laplacian and consequently with homology. 

Accordingly, the method of Rabin provided us with a position space doubling or ``species doubling'' and the conceptual picture was new: the failure in constructing well-defined discrete analogies to the basic operands used in (\ref{dke2}) was  effectively the way doubling survived. Specifically, the Hodge star ($\star$) which is the analogue of $\gamma_5$, and the various properties relating it with the DK operator did not hold simultaneously. The picture is then: i) doubling has an algebraic formulation; and ii) it might be addressed at the classical level, that is at the level of the action.

Meanwhile Becher and Joos attempted to carry out a discrete version of the continuum reduction of the DK equation via the introduction of projection operators and the extraction of the spinorial algebra using representation theory and the Clifford product as multiplication. In the continuum, this leads to the Dirac equation provided one restricts the field to the appropriate subspace $D^{(b)}$; reducing a 16D object to a four component spinor. In fact one can relate DK fermions and staggered fermions as in \cite{Kovacs} for the free case. In matrix form, the reduction is known as ``thinning''.

A relation between field doubling and Geometric Discretisation has been shown in the context of the (topological) abelian Chern-Simons theory \cite{adams}. In the present work \cite{us}, the root discretisation scheme is provided with a Clifford product and a new interpretation is given. We will introduce our method in relation to the two already discussed; first focusing on the algebraic properties and then on the reduction using the Clifford product. The tight relation between the discrete and the continuum theories plays a central role and is described in the last part.

\section{Fermions and Geometry}

Let us start with the DK equation written degree by degree as
\begin{equation}
\label{dke2}
( d \Phi^{(p-1)} - \delta \Phi^{(p+1)} ) = - m \Phi^{(p)}.
\end{equation}
Now, recall the algebraic properties of the co-boundary ($\delta$) and the Hodge star:
\begin{eqnarray}
\label{cob}
\delta&=&\star \, d\, \star, \\
\label{starstar}
\star\star&=& Id. 
\end{eqnarray}
If these two identities were to hold simultaneously in the discrete theory, one could extract a solution of one of the equations in (\ref{dke2}) (that is for given $p$) describing a Dirac field; (\ref{cob}, \ref{starstar}) being used to relate various $p$-degree equations. However, as argued by Rabin this is not possible. 

The geometrical origin is that one is forced to introduce (in our notation) a dual complex $L$ to the original hyper-cubic complex $K$ by subdividing the latter leading to an additional complex which is shifted by half a lattice spacing. In turn, this gives us two discretised versions of the Hodge star exchanging the respective space of cochains (in  $n$-dimensions):
\begin{eqnarray}
\star^{K} : C^{p} (K) \longrightarrow C^{n-p} (L) \\
\star^{L} : C^{p} (L) \longrightarrow C^{n-p} (K)
\end{eqnarray} 
and the induced translation element spoils the identities. Another feature is the way one expresses local chiral transformations in $K$:
\begin{equation}
\Phi^{(4-p)}_{K} \stackrel{\epsilon}{\longmapsto} \Phi^{(4-p)}_{K} + i \epsilon \star^K \Phi^{(p)}_{K}
\end{equation}
and respectively in $L$. The identification of $\gamma_5$ on a spinor field with $\star$ on a differential form is then made. This gives a special meaning to (\ref{starstar}) as the analogue of $\gamma_{5}^{2} = 1$. Not having this property then results in problems with chirality. 

Now let us turn to the approach of Becher and Joos where one needs a discrete Clifford product ($\vee$), an example in the continuum is 
\begin{equation}
dx^{\mu} \vee dx^{\nu} = g^{\mu \nu} + dx^{\mu} \wedge dx^{\nu},
\end{equation}
and with the identification 
\begin{equation}
\gamma_{\mu} = dx^{\mu} \vee (.),
\end{equation}
we immediately obtain the spinorial algebra (with the Euclidean metric):
\begin{equation}
\label{clif}
\{ \gamma^{\mu} , \, \gamma^{\nu} \} = 2 g^{\mu \nu}. 
\end{equation}
The DK equation is found to be invariant under an $SU(4)$ symmetry by right $\vee$-multiplication with a constant differential. Then we use projection operators $P^{(b)}$ to get a new field 
\begin{equation}
\Phi^{(b)} = \Phi \vee P^{(b)}
\end{equation}
which satisfies the Dirac equation expressed as
\begin{equation}
( dx^{\mu} \vee \partial_{\mu} + m ) \Phi^{(b)} = 0.
\end{equation}
The collection of operators $P^{(b)}$ gives rise to the so-called reduction group $\mathcal{R}$. However, the discretised version of this group does not close due to the appearance of translation elements which prevent the reduction to the Dirac equation. In contrast with the former case, they do not originate with the definition of the Hodge star analogue but with the discrete wedge which is used in the definition of the Clifford product. 

The lattice origin of this is the role given to a base point in the definition of the discrete wedge. That is, although the objects are hyper-cubic of any dimensions, the operation of wedging is done at a specific point referred to as a base point.

For example, consider the edge $[01]$ obtained from the vertices $[0]$ and $[1]$ respectively at $x=0$ and $x=h$, then assuming that the base point is set at the vertex $[1]$ for the edge $[01]$, one has to translate the base point from $[1]$ to $[0]$,
\begin{eqnarray}
\label{one}
[01] \wedge [1] &=& [01] \\
\label{two}
[0] \wedge T_{-h}[01] &=& [01]
\end{eqnarray}
This definition is at the origin of the translation elements which plagues the closure of the reduction group in contradiction with the continuum.

From both cases, it appears clearly that the choice of discrete operators is problematic. Let us now move to the new proposal.

\section{The reduction with GD}
In our approach, we map cochains to differential forms in a well-defined way using the Whitney map ($W$). This morphism has the crucial property that it preserves the differential complex namely the relation (de Rham) between the behaviour under $d$ in the continuum and that under the boundary $\partial$ on the chain complex. The map $W$ is then built in the definition of the discrete analogies of the map $( \wedge, \, d, \, \star)$ and for the present purpose, the construction is such that the crucial properties (\ref{cob}, \ref{starstar}) hold. This resolves the first issue. 

The geometric interpretation is as follows: consider the $[01]$ edge again in a square lattice. Then, in the $(x, y)$-plane of the page,
\begin{equation}
\label{whitn}
W([01]) = (1-y) dx 
\end{equation}
The edge is at $y=0$ and the image of the edge has support on the entire square. Next, we define the discrete wedge on the complex $K$ by
\begin{equation}
\label{wedge}
\sigma \wedge^K \eta = A^{K} ( W^{K} ( \sigma) \wedge W^{K} (\eta))
\end{equation}
where the de Rham map $A^K$ plays the role of integration. In the example we take $\sigma = [01]$ and $\eta = [1]$. To return to the question of support, we see that because Whitney elements (of the form  $W ( \sigma)$) have support on the square, no base point is required in order to wedge them. 

The discrete wedge ($\wedge^K$) is then used in the definition of the Clifford product. It is worth noting that ($\wedge^K$) is only associative up to a factor. This does not affect the derivation of a discrete analogue of (\ref{clif}) but when considering the $SU(4)$ invariance of the discrete DK equation, one avoids ``moving the brackets'' to establish the discrete invariance \cite{us}. In \cite{BJ} one use a special case where cochains of specific degrees are considered. 

In summary, the discrete theory spans a space of functions which gives the desired support and yet the theory is local, since a field such as (\ref{whitn}) has non-zero value solely in each and every square cell having $[01]$ as a boundary component.

Again, we are using a doubling of spaces, namely $K$ and $L$. While because of the issue of support, the $\star$ has no shift and (up to signs)
\begin{eqnarray}
\star^K \star^L &= Id^L, \;\; \delta^K = \partial^K = \star^L d^L \star^K, \\
\star^L \star^K &= Id^K, \;\; \delta^L = \partial^L = \star^K d^K \star^L.
\end{eqnarray}
We then have the usual relations in $K$ say,
\begin{eqnarray}
\{ \gamma_5, \, D \}&=&\star^{K}(d^K - \delta^K ) + (d^L - \delta^L )\star^K \\
&=& 0 
\end{eqnarray}
involving the two spaces $K$ and $L$. This brings us to the action functional, it is based on the natural inner product (used to define the $\star$):
\begin{equation}
\label{inner}
< \sigma, \, \eta> = \int_{M} W^B (B \sigma ) \wedge W^B ( B \eta^K ), 
\end{equation}
where $B$ is the finer complex which is the union of $K$ and $L$.
Hence, given the field in the sublattice $K$, it has a representative in $L$ given by $\star^K \Phi^K$ which is itself the image of a field in $K$, we then have an action for $K$-fields:
\begin{equation}
S_K = < \bar{ \Phi}_K, ( d^K - \delta^K + m ) \Phi_K> 
\end{equation}
The pairing between the two spaces $K$ and $L$ is interpreted as an inner product and does not amount to a coupling of independent $K$ and $L$ fields. In turn we define a direct sum of independent actions $S_K$ and $S_L$. Then, the partition function splits into two components, one for $C(K)$-fields and the other for $C(L)$-fields.
So, the use of the inner product (\ref{inner}) is the more natural way of defining the action functional which also has the virtue of avoiding the introduction of base points.

Finally a comment about gauging. By retaining inhomogeneous cochains, one has to be careful when gauging. This step when completed should shed some light on the applications of the new technique. General features are, i) the choice of minimal coupling in the adjoint representation and ii) the insertion of a parallel transport operator $U$ in the action on the left of the DK operator and does not conflict with the reduction. In the square lattice under consideration, obvious discrete symmetries of the action are present.  
\section{Conclusion}
By working in the DK picture and using the cochain space in a way that relates closely to the continuum fermion theory, one avoids the ``lattice'' hypothesis of the Nielsen-Ninomiya theorem where the fermions are defined at vertices. This new technique describes a finite theory and careful analysis of the gauging remains to be done.


\begin{thebibliography}{9}
\bibitem{us}
V. de Beauc\'e and S. Sen, hep-th/0305125
\bibitem{Us}
S. ~Sen, Siddhartha ~Sen, J.~C.~Sexton and D.~H.~Adams,
{\em Phys. Rev.} {\bf E61}, 3174 (2000) 
\bibitem{NN}
H.~ B.~ Nielsen and M.~ Ninomiya {\bf B105} 219 (1981).
\bibitem{rabin}
J.~M.~Rabin, {\em Nuclear Physics} {\bf B201}, 315 (1982).
\bibitem{BJ}
P.~Becher and H.~Joos, {\em Z.~Phys.} {\bf C15}, 343 (1982).
\bibitem{Kovacs}
G.~T~Bodwin, E.~V~Kov\'acs, {\em Phys. Rev.} {\bf D38}, 1206 (1988).
\bibitem{adams}
D. H ~Adams, {\em Phys. Rev. Lett.} {\bf 78}, 4155-4158,  (1997).
\end{thebibliography}
\end{document}